\documentstyle[aps,epsfig]{revtex}
\begin{document}
\draft
\title{TWO-PHOTON PRODUCTION OF FOUR-QUARK STATES\\ UP TO THE $J/\psi$ ENERGY
\thanks{Talk presented by G.N. Shestakov at the International Workshop on $e^+e
^-$ Collisions from $\phi$ to $J/\psi$, Budker Institute of Nuclear Physics, 
Novosibiksk, Russia, March 1-5, 1999.}}
\author{N.N. Achasov and G.N. Shestakov\\[0.3cm]}
\address{Laboratory of Theoretical Physics, S.L. Sobolev Institute for 
Mathematics,\\ 630090, Novosibirsk 90, Russia}
\maketitle
\begin{abstract} Evidence for an explicitly exotic state with isospin 2 and
spin-parity $2^+$ near the $\rho\rho$ threshold  and nontrivial complementary
indications of the unusual quark composition of the $f_0(980)$ and $a_0(980)$
states obtained from the reactions of two-photon formation of neutral meson
resonances are discussed, together with puzzling phenomena in the channels
$\gamma\gamma\to\rho^0\phi$ and $\gamma\gamma\to\rho^0\rho^0$ at high energies.
\end{abstract} 
\begin{flushleft}{\bf 1. \ INTRODUCTION}\end{flushleft}

One of the most striking effects found in two-photon reactions is the large
cross section for the reaction $\gamma\gamma\to\rho^0\rho^0$ near its threshold
and at the same time the much lower cross section for the reaction $\gamma
\gamma\to\rho^+\rho^-$ that rules out an ordinary $q\bar q$ resonance
interpretation (for reviews see e.g. [1-3]). Another important fact is the
smallness of the two-photon widths for $f_0(980)$
and $a_0(980)$ resonances. All these phenomena speaks about manifestations of
four-quark dynamics. Here we first discuss briefly the current situation on the
reactions $\gamma\gamma\to\rho\rho$ at the $\gamma\gamma$ centre-of-mass energy
$(W_{\gamma\gamma})$ near the nominal $\rho\rho$ threshold. Then we give a
short overview of the available results on the two-photon widths of the $f_0(98
0)$ and $a_0(980)$ states, together with comments on the reaction $\gamma\gamma
\to K^+K^-$ and on the inverted mass spectrum of the light scalar nonet in
the context of SU(3). In addition, we pay attention to the probable strong
violation of the conventional factorized Pomeron exchange model in the
reactions $\gamma\gamma\to\rho^0\phi$ and $\gamma\gamma\to\rho^0\rho^0$ at high
energies.\begin{flushleft}{\bf 2. \ ON THE EXOTIC $X(I^G(J^{PC})=2^+(2^{++}))$
STATE}\end{flushleft}

Since 1992 the Particle Data Group has quoted two non-$q\bar q$ candidates in
the resonance states with explicitly exotic quantum numbers: the $\hat\rho(1405
)$ with $I^G(J^{PC})=1^-(1^{-+})$ is from hadroproduction [4] and $X(1600)$
with $I^G(J^{PC})=2^+(2^{++})$ is from the reactions $\gamma\gamma\to\rho\rho$
[1-4]. Hitherto, the most clear evidence for the presence of the $X(1600)$ in
the $\rho\rho$ channels has been obtained by the ARGUS Collaboration in two
high statistics experiments [5,6]. Their results are shown in Fig. 1. The
observed difference between the $\rho^0\rho^0$ and $\rho^+\rho^-$ partial cross
sections with $(J^P,\,|J_z|)=(2^+,\,2)$ (where $J_z$ is the helicity of the
intermediate state $J^P$) can be naturally explained by the hypothetical $X(1600
)$ state contribution. There are the following isotopic relations between the
reaction amplitudes and amplitudes with definite isospin: $A(\gamma\gamma\to\rho
^0\rho^0)=(A_{I=0}+2A_{I=2})/(3\sqrt{2})$ and $A(\gamma\gamma\to\rho^+\rho^-)=(
A_{I=0}-A_{I=2})/3$, where the identity of $\rho^0$ mesons is considered in the
normalization of the amplitude $A(\gamma\gamma\to\rho^0\rho^0)$. Thus, for an
ordinary isospin 0 resonance one expects $\sigma(\gamma\gamma\to\rho^+\rho^-)/
\sigma(\gamma\gamma\to\rho^0\rho^0)=2$ and for a pure isospin 2 resonance
$\sigma(\gamma\gamma\to\rho^+\rho^-)/\sigma(\gamma\gamma\to\rho^0\rho^0)=1/2.$
Instead the observed ratio is lower than 1/2. A resonance interpretation
for such a result in terms of $q^2\bar q^2$ states thus requires the presence
of a flavor exotic $I=2$ resonance which interferes with some isoscalar
contributions. Such a distinct manifestation of the tensor four-quark state
with $I=2$ in the reactions $\gamma\gamma\to\rho\rho$ was predicted [7,8] on
the basis of the MIT bag model [9].

Similar to the other candidates in ``certified" exotic states, the state $X^0(1
600,\,2^+(2^{++}))$ is in need of further confirmations. So, its doubly charged
partners could be looked for in hadroproduction, for example, in the reactions 
$\pi^+p\to X^{++}n\to\rho^+\rho^+n$, $\pi^-p\to X^{--}\Delta^{++}\to\rho^-\rho
^-\Delta^{++}$, and $pp\to n(X^{++})n\to n(\rho^+\rho^+)n$ [10]. Recently we 
have also shown that the search for $X^{+}$ and $X^{-}$ states is quite
feasible in the photoproduction reactions $\gamma N\to X^\pm N\to\rho^\pm\rho^0
N$ and $\gamma N\to X^\pm\Delta\to\rho^\pm\rho^0\Delta$ with the help of the
intensive 6 GeV photon beam at Jefferson Laboratory [11]. The expected yield of
the $\gamma N\to X^\pm N\to\rho^\pm\rho^0N$ events in a 30-day run approximates
$2.8\times10^6$. This estimate should be compared with 16000 events collected
for $\gamma\gamma\to\pi^+\pi^-\pi^+\pi^-$ by the TASSO, MARK II, CELLO, PLUTO,
TPC/2$\gamma$, and ARGUS Collaborations [2]. 

Let us note that the L3 Collaboration at LEP-2 recently
begun the second stage of the examination of the reactions $
\gamma\gamma\to VV'$ ($V(V')=\rho,\omega,\phi,K^*$) with higher statistics, and
very interesting results on the reaction $\gamma\gamma\to\pi^+\pi^-\pi^+\pi^-$
for $0.75\leq W_{\gamma\gamma}\leq4.9$ GeV have been presented at this Workshop
[12]. For $W_{\gamma\gamma}<2$ GeV, $\sigma(\gamma\gamma\to\pi^+\pi^-\pi^+\pi^-
)$ is strongly dominated by $\rho^0\rho^0$ production [12] and is in good
agreement with the ARGUS data on $\gamma\gamma\to\rho^0\rho^0$ shown in Fig. 1.
\begin{figure}\centerline{\epsfig{file=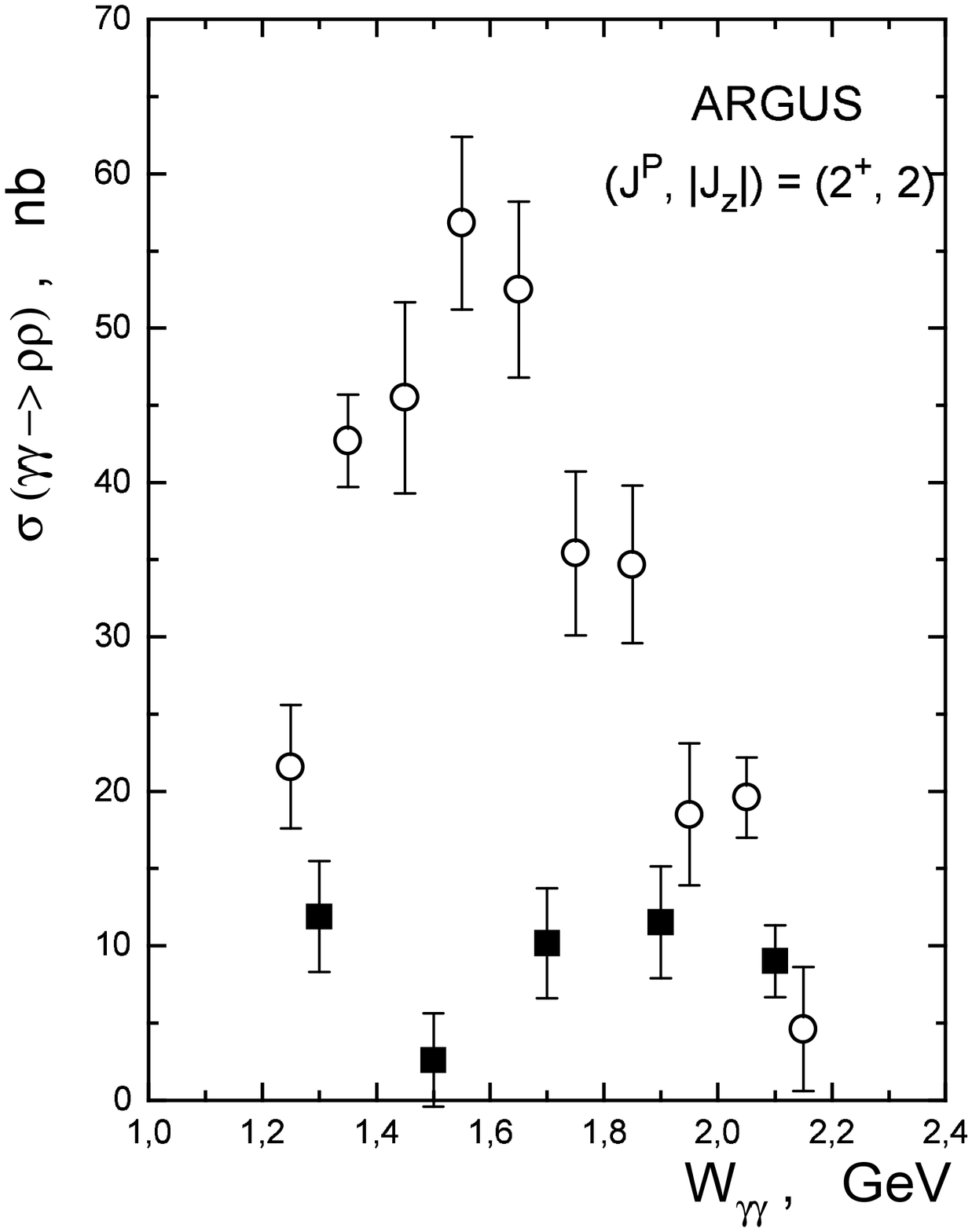, height=14.5cm, width=10cm}}
\caption{The ARGUS data on the partial cross sections for $\gamma\gamma\to
\rho^0\rho^0$ [5] (open circles) and $\gamma\gamma\to\rho^+\rho^-$ [6] (full
squares) with $\,(J^P,|J_z|)=(2^+,2)\,$.}\end{figure}\vspace*{0.2cm}

\begin{flushleft}{\bf 3. \ TWO-PHOTON WIDTHS OF THE $f_0(980)$ AND $a_0(980)$
MESONS}\end{flushleft}

As is well known the reaction $\gamma\gamma\to\pi^+\pi^-$ for $W_{\gamma\gamma}
<1.5$ GeV is dominated by the Born term and $f_2(1270)$ resonance
contributions,
and only some wiggle-waggle is observed in the $f_0(980)$ resonance region.
Similarly only some small enhancement due to $f_0(980)$ production has been
found in $\gamma\gamma\to\pi^0\pi^0$. A more clear signal from the $a_0(980)$
is observed in the reaction $\gamma\gamma\to\pi^0\eta$ owing to the suppression
of $a_2(1320)$ production in the $\pi^0\eta$ channel. The existing experimental
results [13-17] on the two-photon width of the $f_0(980)$ are listed in the
upper part of Table 1. However, the Particle Data Group [4] has ignored, for
unknown reasons, all these results except the JADE data. Combining these
data with the earlier analysis performed by Morgan and Pennington [18], they
quote a distinctly overestimated value of $0.56\pm0.11$ keV for the $f_0(980)
\to\gamma\gamma$ decay width (see the middle part of Table 1). Fortunately,
very recently Boglione and Pennington have performed a new analysis [19] and
found a much smaller value of $0.28+0.09-0.13$ keV (see Table 1). The data for
the two-photon width of the $a_0(980)$ meson [4,15,20] are listed in Table 2.

\begin{center}\begin{tabular}{|ll|}
\multicolumn{2}{l}{{\bf Table 1:} \ $\Gamma(f_0(980)\to\gamma\gamma)$ in keV}
\\[0.09cm] \hline
$\ \ 0.31\pm0.14\pm0.09\ \ $ & \ \ \ \  CBALL (90)\ [13]\ \   \\
$\ \ 0.29\pm0.07\pm0.12\ \ $ & \ \ \ \  MARK II (90)\ [14]\ \   \\
$\ \ 0.42\pm0.06+0.08-0.18\ \ $ & \ \ \ \ JADE (90)\ [15]\ \ \\
$\ \ 0.25\pm0.10\ \ $ &\ \ \ \ CBALL, Karch (91)\ [16]\ \ \\
$\ \ 0.20\pm0.07\pm0.04\ \ $ & \ \ \ \ CBALL, Bienlein (92)\ [17]\ \ \\
$\ \ \leq0.31\ \ (90\%$ CL)\ \ & \ \ \ \ CBALL, Bienlein (92)\ [17]\ \ \\
\hline $\ \ 0.56\pm0.11\ \ $ & \ \ \ \ PDG (98)\ [4]\ \ \\
$\ \ 0.42\pm0.06\pm0.18\ \ $ & \ \ \ \ JADE (90)\ [4,15]\ \ \\
$\ \ 0.63\pm0.14\ \ $ & \ \ \ \ Morgan, Pennington (90)\ [18]\ \ \\ \hline
$\ \ 0.28+0.09-0.13\ \ $ & \ \ \ \ Boglione, Pennington (98)\ [19]\ \
\\ \hline\multicolumn{2}{c}{}\\
\multicolumn{2}{l}{{\bf Table 2:} \ $\Gamma(a_0(980)\to\gamma\gamma)BR(a
_0(980)\to\pi\eta)$ in keV}\\[0.09cm]\hline
$\ \ 0.24+0.08-0.07\ \ \qquad$ & \ \ \ \ PDG (98)\ [4]\ \ \\
$\ \ 0.19\pm0.07+0.10-0.07\ \ \qquad$ & \ \ \ \ CBALL (86)\ [20]\ \ \\
$\ \ 0.28\pm0.04\pm010\ \ \qquad$ & \ \ \ \ JADE (90)\ [15]\ \ \\
\hline\end{tabular}\end{center}
\vspace*{0.2cm}

All these results should be compared with the well known two-photon widths of
the tensor mesons [4], $\Gamma(f_2(1270)\to\gamma\gamma)=2.8\pm0.4\,$keV and $
\Gamma(a_2(1320)\to\gamma\gamma)=1.00\pm0.06\,$keV, and also with the following
relations predicted by the $q\bar q$ model (see e.g. [19]): $\Gamma(f_2
\to\gamma\gamma):\Gamma(a_2\to\gamma\gamma):\Gamma(f_2'\to\gamma\gamma)=25: 9:
2$ and $\Gamma(0^+\to\gamma\gamma)=(15/4)\times\Gamma(2^+\to\gamma\gamma)\times
(m_{0^+}/m_{2^+})^3$. Hence it follows, for example, that $\Gamma(a_0(980)\to
\gamma\gamma)\approx1.6\,$keV. That is too much. On the other hand, the
four-quark scheme gives $\Gamma(f_0(980)\to\gamma\gamma)\approx\Gamma(a_0(980)
\to\gamma
\gamma)\approx0.27\,$keV [2,7,21]. This tentative estimate is in reasonable
agreement with the current experimental situation, which clearly speaks in
favour of the unusual structure of the $f_0(980)$ and $a_0(980)$ resonances.
Certainly, the two-photon widths are the nonunique indication of such a kind.
For example, in contrast to the reaction $\gamma
\gamma\to\pi^+\pi^-$, there are not any signs of the expected huge S-wave Born
term contribution near the threshold of the reaction $\gamma\gamma\to
K^+K^-$ [1]. The reduction of the Born term in $\gamma\gamma\to K^+K^-$ can
be explained by the resonant $K^+K^-$ final state interaction due to the $f_0(9
80)$ and $a_0(980)$ resonances [22]. It should be emphasized, firstly, that
such a reduction is the straightforward consequence of the unitarity condition
and, secondly, that it is really possible only if the $f_0(980)$ and $a_0(980
)$ are strongly coupled to the $K\bar K$ channels, for instance, as in the
four-quark model.

At present the problem of scalar mesons is considered in many aspects and
there are much evidences that the $f_0(980)$ and $a_0(980)$ states involve
four quarks [23]. Let us, for example, look at the mass spectrum of the light
scalar nonet [$\sigma(600)$, $\kappa(900)$, $a_0(980)$, $f_0(980)$], which
currently is the subject of wide speculation (see e.g. [9,24-26]). It is
obvious that this spectrum is inverted in comparison with those of
the light vector and tensor nonets (see the following diagrams, where the state
masses increase from bottom to top).\vspace*{0.07cm}
\begin{center}\begin{tabular}{ccc}
$\ a^+_0\qquad a^0_0\,/\,f_0\qquad a^-_0$\ \ \ \ & $\ \ \phi$ & $\ \ \ f'_2$
\\[15pt]$\kappa\ \ \qquad\ \ \bar\kappa\ \ $ & $\ \ \ K^*\ \qquad\ \bar K^*$
& $\ \ \ \ K^*_2\ \qquad\ \bar K^*_2$ \\[15pt] $\sigma\ \ \ $ & $\ \ \ \ \ \rho
^+\qquad \rho^0\,/\,\omega\qquad \rho^-$\ \ \ \ & \ \ \ \ \ $a^+_2\qquad a^0_2
\,/\,f_2\qquad a^-_2$\end{tabular}\end{center}\vspace*{0.1cm}
However, within the framework of SU(3)-symmetry (but not in the $q\bar
q$ model), the Gell-Mann -- Okubo mass formula for an ideal mixed nonet, $$4
\,M^2_{I=1/2}\,=\,M^2_{I=1}+2\,M^2_{I=0'}+M^2_{I=0}\ ,\qquad M^2_{I=1}=M^2_{I=0
}\ ,$$ has two solutions. Solution I is
$$4(m^2_0+\Delta)_{I=1/2}\,=\, (m^2_0)_{I=1}+2(m^2_0+2\Delta)_{I=0'}+(m^2_0)_{I
=0}\ ,$$ and solution II $$4(m^2_0+\Delta)_{I=
1/2}\,=\,(m^2_0+2\Delta)_{I=1}+2(m^2_0)_{I=0'}+(m^2_0+2\Delta)_{I=0}\ ,$$
where $\Delta$ is due to SU(3) breaking. Furthermore, the system of the SU(3)
relations between the coupling constants for the ideal nonet members also has
two solutions compatible with the Okubo-Zweig-Iizuka (OZI) rule. Solution I
gives that the isoscalar undegenerate with the isovector uncouples to $\pi\pi$.
That is, for example for the usual tensor nonet, we have the following main
decays: $$\ f'_2\to K\bar K\,,\ \ \ \ a_2\to\pi\eta\,,\ K\bar K\,,\ \ \ \
f_2\to\pi\pi\,,\ K\bar K\,.$$ Solution II gives that the isoscalar
degenerate with the isovector uncouples to $\pi\pi$ (also there arises an
extraordinary prediction that the lighter isoscalar uncouples to $K\bar K$).
Applying this solution to the light scalar nonet, we have the transitions $$\
\sigma\to\pi\pi\,,\ \ \ \ a_0\to\pi\eta\,,\ K\bar K\,,\ \ \ \ f_0\to K\bar
K\,.$$ Thus, there is no problem of the inverted nonet in the context of SU(3).
Within the quark model, solution I for the masses and coupling constants
corresponds to the conventional $q\bar q$ states, whereas solution II most of
all corresponds to the $q^2\bar q^2$ states decaying (if a phase space permits)
by the OZI-superallowed way [9]. In particular, it is seen that the degenerate
four-quark states $a_0$ and $f_0$ contain strange quarks and both strongly
couple to the $K\bar K$ channels.
\begin{flushleft}{\bf 4. \ PUZZLE OF THE REACTIONS $\gamma\gamma\to\rho^0\phi$
AND $\gamma\gamma\to\rho^0\rho^0$}\end{flushleft}

According the ARGUS data [27] and the new data from the L3 Collaboration [12]
$\sigma(\gamma\gamma\to\rho^0\phi)=0.16\pm0.16\,$nb for $3.25\leq W_{\gamma
\gamma}\leq3.5$ GeV and $\sigma(\gamma\gamma\to\rho^0\rho^0)<1.5\,$nb for $4.
5\leq W_{\gamma\gamma}\leq4.9$ GeV respectively. At high energies the $\rho^0
\phi$ and $\rho^0\rho^0$ production cross sections can be estimated by using
the factorized Pomeron exchange model. In the $W_{\gamma\gamma}$ region from
11.5 to 18.4 GeV, such estimates yield $\sigma(\gamma\gamma\to\rho^0\phi)=1.2-2
.4\,$nb and $\sigma(\gamma\gamma\to\rho^0\rho^0)=9.9-21\,$nb (for details see
[28]).
Hence, in the range between the maximal reached energies and $W_{\gamma\gamma}
\approx11.5$ GeV the $\gamma\gamma\to\rho^0\phi$ and $\gamma\gamma\to\rho^0\rho
^0$ cross sections can increase by an order of magnitude. Nothing of the kind
has yet happened in elastic and quasielastic reactions with the Pomeron
exchange and with particles involving light quarks. An unusually strong rise of
$\sigma(\gamma\gamma\to\rho^0\phi)$ and $\sigma(\gamma\gamma\to\rho^0\rho^0)$
would be a real challenge for our current knowledge about the dynamics of
quasi-two-body reactions. Here either we face a new physical phenomenon or the
ARGUS and L3 data have been underestimated for some reason. However, the latter
possibility seems almost improbable. Moreover, if the two cross sections do not
increase approximately by an order of magnitude up to $W_{\gamma\gamma}\approx
11.5$ GeV, then it will speak about the strong failure of the conventional
factorization model in the reactions $\gamma\gamma\to\rho^0\rho^0$ and $\gamma
\gamma\to\rho^0\phi$ in the energy region where this works fairly well in all
other cases [28]. Thus, direct measurements of $\sigma(\gamma\gamma\to\rho
^0\phi)$ for $W_{\gamma\gamma}>3.5$ GeV and $\sigma(\gamma\gamma\to\rho^0\rho^0
)$ for $W_{\gamma\gamma}>4.9$ GeV (and the cross sections for other reactions $
\gamma\gamma\to VV'$ at high energies) would be very desirable.
\begin{flushleft}{\bf 5. \ CONCLUSION}\end{flushleft}

There are quite a number of important issues in hadrodynamics which can be
elucidated using two-photon reactions. It would be very important to define
more precisely $\sigma(\gamma\gamma\to a_0(980)\to\pi^0\eta)$ and $\Gamma(a_0(9
80)\to\gamma\gamma)$, $\,\sigma(\gamma\gamma\to f_0(980)\to\pi^0\eta)$ and $
\Gamma(f_0(980)\to\gamma\gamma)$, and also the S-wave partial cross section for
the reaction $\gamma\gamma\to K^+K^-$ near the thresholds. Moreover, the second
stage of the high statistics investigations of the reactions $\gamma\gamma\to 
VV'$ can become a crucial test for the four-quark states from the MIT-bag, in
particular, for the explicitly exotic state $X(1600)$.

Finally, it would be very interesting to know from the L3 Collaboration whether
there is the strong violation of the conventional factorized Pomeron exchange 
model in the reactions $\gamma\gamma\to VV'$ at high energies.
\begin{flushleft}{\bf REFERENCES}\end{flushleft}
\begin{enumerate}
\item M. Poppe, Intern. J. Mod. Phus. A1 (1986) 545.\\ Ch. Berger and W.
Wagner, Phys. Rep. 146 (1987) 1.\\ M. Feindt and J. Harjes, Nucl. Phys. B
(Proc. Suppl.) 21 (1991) 61.
\item N.N. Achasov and G.N. Shestakov, Usp. Fiz. Nauk 161 (1991) No.6, 53
[Sov. Phys. Usp. 34 (1991) No.6, 471].
\item H. Albrecht et al., Phys. Rep. 276 (1996) 223.
\item C. Caso et al. (Particle Data Group), Euro. Phys. J. C3 (1998) 1.
\item H. Albrecht et al., Z. Phys. C50 (1991) 1.
\item H. Albrecht et al., Phys. Lett. B267 (1991) 535.
\item N.N. Achasov, S.A. Devyanin, and G.N. Shestakov, Phys. Lett. 108B (1982)
134; Z. Phys. C16 (1982) 55; ibid. C27 (1985) 99.
\item B.A. Li and K.F. Liu, Phys. Lett. 118B (1982) 435 and Er. ibid. 124B
(1983) 550; Phys. Rev. Lett. 51 (1983) 1510.
\item R.L. Jaffe, Phys. Rev. D15 (1977) 267, 281.
\item N.N. Achasov and G.N. Shestakov, Intern. J. Mod. Phys. A7 (1992) 4313.
\item N.N. Achasov and G.N. Shestakov, hep-ph/9901380, 1999.
\item V. Schegelsky, these proceedings.
\item H. Marsiske et al., Phys. Rev. D41 (1990) 3324.
\item J. Boyer et al., Phys. Rev. D42 (1990) 1350.
\item T. Oest et al., Z. Phys. C47 (1990) 343.
\item K.-H. Karch, Procs. XXVth Recontre de Moriond, Les Arcs, France, 1991,
ed. J. Tr\^anh Thanh V\^an, Editions Frontieres, 1991, p. 423.
\item J.K. Bienlein, Procs. IXth Intern. Workshop on Photon-Photon Collisions,
San Diego, 1992, ed. D. Caldwell and H.P. Paar, World Scientific, 1992, p. 241.
\item D. Morgan and M.R. Pennington, Z. Phys. C48 (1990) 623.
\item M. Boglione and M.R. Pennington, hep-ph/9812258, 1998.
\item D. Antreasyan et al., Phys. Rev. D33 (1986) 1847.
\item N.N. Achasov and G.N. Shestakov, Z. Phys. C41 (1988) 309.
\item N.N. Achasov and G.N. Shestakov, Mod. Phys. Lett. A9 (1994) 1351.
\item N.N. Achasov, Usp. Fiz. Nauk 168 (1998) 1257; these proceedings.
\item N.N. Achasov, S.A. Devyanin, and G.N. Shestakov, Usp. Fiz. Nauk 142
(1984) 361 [Sov. Phys. Usp. 27 (1984) 161].\\
N.N. Achasov and G.N. Shestakov, Phys. Rev. D49 (1994) 5779.
\item D. Black, A.H. Fariborz, F. Sannino, and J. Schechter, Phys. Rev. D59
(1999) 074026.
\item M. Ishida, Prog. Theor. Phys. 98 (1999) No.3; hep-ph/9902260, 1999.
\item H. Albrecht et al., Phys. Lett. B332 (1994) 451.
\item N.N. Achasov and G.N. Shestakov, Phys. Rev. D52 (1995) 6291;
hep-ph/9903409, 1999.\end{enumerate}\end{document}